\documentclass[aps,prd,twoside,twocolumn,nofootinbib,showpacs,floatfix]{revtex4-1}
\usepackage{amssymb}
\usepackage{amsmath,amssymb}
\usepackage{graphicx,bm}
\usepackage{slashed}
\usepackage{times}
\usepackage{epstopdf}
\usepackage{ulem} 
\usepackage[usenames]{color}
\usepackage{float}
\usepackage{subfigure}
\usepackage{subfigure}
\usepackage{rotating}
\usepackage{color}
\usepackage{multirow}
\usepackage{dcolumn}
\usepackage{overpic}
\usepackage{booktabs}
\usepackage{appendix}
\usepackage{makecell}
\usepackage{diagbox}

\renewcommand\sout{\bgroup \color{red} \ULdepth=-.5ex \ULset}

\usepackage[colorlinks, citecolor=blue,anchorcolor=red,menucolor=red,linkcolor=blue,filecolor=red,runcolor=red,urlcolor=blue,frenchlinks=red]{hyperref}
\begin{document}
\title{SU(3) breaking effect in the $Z_c$ and $Z_{cs}$ states}

\author{Kan Chen$^{1,2,3,4,5}$}\email{chenk10@nwu.edu.cn}

\affiliation{
$^1$School of Physics, Northwest University, Xi'an 710127, China\\
$^2$Shaanxi Key Laboratory for theoretical Physics Frontiers, Xi'an 710127, China\\
$^3$Institute of Modern Physics, Northwest University, Xi'an 710127, China\\
$^4$Peng Huanwu Center for Fundamental Theory, Xi'an 710127, China\\
$^5$School of Physics and Center of High Energy Physics, Peking University, Beijing 100871, China}
\begin{abstract}
Based on the hadronic resonance picture, we propose a possible framework to simultaneously describe the resonance parameters of the observed $Z_{cs}$ and $Z_{c}$ states that are close to the thresholds of the $D^{(*)}\bar{D}^{(*)}_s$ and $D^{(*)}\bar{D}^{(*)}$ systems, respectively. We construct the effective potentials of the $Z_{cs}$ and $Z_{c}$ states by analogy with the effective potentials of the leading order (LO) and next-to-leading-order (NLO) $N-\bar{N}$ interactions. Then we introduce an SU(3) breaking factor $g_x$ to identify the differences between the effective potentials in the $Z_{cs}$ and $Z_{c}$ states. We perform two calculations to discuss the differences and similarities of the $Z_{cs}$ and $Z_{c}$ states. In the first calculation, we adopt the LECs extracted from the experimental $Z_{cs}$ states to calculate the $Z_c$ states, and show that if the $Z_{cs}(4000)$ is the SU(3) partner of the $Z_c(3900)$, then our framework can reproduce the large width difference between the $Z_{cs}(4000)$ and $Z_{c}(3900)$ by adjusting the SU(3) breaking factor in a reasonable region. Besides, this SU(3) breaking effect also accounts for the absence of a $Z_{c}$ state with $J^{PC}=1^{++}$, which should be the SU(3) partner of the $Z_{cs}(3985)$. In the second calculation, we separately fit the LECs from the experimental $Z_{cs}$ and $Z_{c}$ states. We show that these two sets of LECs are very similar to each other, indicating a unified set of LECs that could describe the effective potentials of the $Z_{cs}$ and $Z_{c}$ states simultaneously. Then we proceed to systematically predict the other possible $Z_{cs}$ and $Z_{c}$ states that are close to the $D^{(*)}\bar{D}^{(*)}_s$ and $D^{(*)}\bar{D}^{(*)}$ systems, respectively. Further explorations on the $Z_{cs}$ states would be crucial to test our theory.
\end{abstract}
\maketitle
\section{Introduction}\label{Sec 1}
In recent years, several charged hidden charm states that are close to the thresholds of the $D^{(*)}\bar{D}^{(*)}_s$ and $D^{(*)}\bar{D}^{(*)}$ systems are discovered in various experiments \cite{ParticleDataGroup:2022pth,BESIII:2020qkh,BESIII:2022qzr,LHCb:2021uow,LHCb:2023hxg,BESIII:2022vxd,LHCb:2021uow,Belle:2008qeq,Belle:2014wyt,LHCb:2018oeg}. In Table \ref{Exp}, we list their masses, widthes and observed channels. These states have exotic quantum numbers, the identifications of their exotic natures are straightforward. Their possible underlying structures are extensively discussed in many literatures (see reviews \cite{Chen:2016qju,Lebed:2016hpi,Esposito:2016noz,Hosaka:2016pey,Guo:2017jvc,Ali:2017jda,Liu:2019zoy,Brambilla:2019esw,Lucha:2021mwx,Chen:2021ftn,Chen:2022asf,Meng:2022ozq}).


\begin{table}[htbp]
\renewcommand\arraystretch{1.3}
\caption{The observed $Z_{cs}$ and $Z_{c}$ states that are close to the $D^{(*)}\bar{D}^{(*)}_s$ and $D^{(*)}\bar{D}^{(*)}$ thresholds, respectively. The masses and widthes are in units of MeV. The statistical and systematical errors have been added in quadrature.\label{Exp}}
\setlength{\tabcolsep}{0.0mm}{
\begin{tabular}{cccccccc}
\toprule[1pt]
State&$I(J^P)$&Mass&Width&Decay channel\\
\hline
$Z_c(3900)^{\pm}$\cite{ParticleDataGroup:2022pth}&$1^+(1^{+-})$&$3887\pm3$&$28\pm3$&$J/\Psi \pi$, $D\bar{D}^*$\\
$Z_c(4020)^{\pm}$\cite{ParticleDataGroup:2022pth}&$1^{+}(?^{?-})$&$4024\pm2$&$13\pm5$&$h_c(1P)\pi$, $D^*\bar{D}^*$\\
$Z_c(4050)^{\pm}$\cite{Belle:2008qeq}&$1^{-}(?^{?+})$&$4051^{+24}_{-43}$&$82^{+52}_{-28}$&$\pi^+\chi_{c1}(1P)$\\
$Z_{c}(4055)^{\pm}$\cite{Belle:2014wyt}&$1^{+}(?^{?-})$&$4054\pm3$&$45\pm13$&$\pi^+\Psi(2S)$\\
$Z_{c}(4100)^{\pm}$\cite{LHCb:2018oeg}&$1^-(?^{??})$&$4096^{+27}_{-30}$&$152^{+84}_{-68}$&$\eta_c(1S)\pi^-$\\
\hline
State&$I^{G}(J^{PC})$&Mass&Width&Decay channel\\
\hline
$Z_{cs}(3985)^+$\cite{BESIII:2020qkh,BESIII:2022qzr}&$\frac{1}{2}(?^{?})$&$3985\pm3$&$14^{+10}_{-7}$&$D_s^+\bar{D}^{*0}+D_s^{*+}\bar{D}^0$ \\
$Z_{cs}(4000)^+$\cite{LHCb:2021uow,LHCb:2023hxg}&$\frac{1}{2}(1^+)$&$4003^{+7}_{-15}$&$131\pm30$&$J/\Psi K^+$\\
$Z_{cs}(4123)^{-}$\cite{BESIII:2022vxd}&$\frac{1}{2}(?^?)$&$4124\pm5$&$-$&$D_s^{*-}D^{*0}+c.c.$\\
$Z_{cs}(4220)^+$\cite{LHCb:2021uow}&$\frac{1}{2}(1^+)$&$4216^{+49}_{-38}$&$233^{+110}_{-90}$&$J/\Psi K^+$\\
\bottomrule[1pt]
\end{tabular}}
\end{table}

In the $Z_c$ sector, the spin-parity number of $Z_c(3900)$ is measured to be $1^+$ \cite{BESIII:2017bua}. The spin-parity number of $Z_{c}(4020)$ \cite{ParticleDataGroup:2022pth} is not measured yet, but since the masses of $Z_c(3900)$ and $Z_c(4020)$ are above the thresholds of the $D\bar{D}^*$ and $D^*\bar{D}$ by a few MeV, respectively, and they both have narrow widthes. Thus, the $Z_c(4020)$ is assumed to be the  heavy quark spin (HQS) partner of the $Z_c(3900)$ and have $J^{P}=1^{+}$. 

The states $Z_c(4050)$ \cite{Belle:2008qeq}, $Z_{c}(4055)$ \cite{Belle:2014wyt}, and $Z_c(4100)$ \cite{LHCb:2018oeg} listed in Table \ref{Exp} still need further confirmation. Among them, the $Z_c(4055)$ is reported \cite{Belle:2014wyt} in the $e^+e^-\rightarrow \pi^+\pi^-\Psi(2S)$ via ISR process. Although the resonance parameters of the $Z_{c}(4055)$ are extracted to be $M=4054\pm3\pm1$ MeV and $\Gamma=45\pm11\pm6$ MeV. However, after taking into account the interference effect between the $\pi^+\pi^-$ amplitude and the $Z_c$ amplitude, further preliminary PWA analysis \cite{Bondar:2018} shows that the resonance parameters of this structure could become $M=4019.0\pm1.9$ MeV and $\Gamma=29\pm4$ MeV, which are consistent with the resonance parameters of $Z_c(4020)$. If such PWA analysis is confirmed, the $Z_c(4020)$ can also decay into $\pi\Psi(2S)$ final states. The $Z_{c}(4100)^{\pm}$ \cite{LHCb:2018oeg} is reported in the $B^0\rightarrow K^+\pi^-\eta_c$ process, the spin-parity assignments $J^P=1^{-}$ and $0^+$ are both consistent with the data. Besides, the $Z_c(4050)^{\pm}$ is reported in the $B^0\rightarrow K^+\pi^-\chi_{c1}$ process \cite{Belle:2008qeq}. The possible interpretations to the above $Z_c$ states include the molecular states, tetraquark states, and kinematical effects (see reviews \cite{Chen:2016qju,Lebed:2016hpi,Esposito:2016noz,Hosaka:2016pey,Guo:2017jvc,Ali:2017jda,Liu:2019zoy,Brambilla:2019esw,Lucha:2021mwx,Chen:2021ftn,Chen:2022asf,Meng:2022ozq}).

In the $Z_{cs}$ sector, the $Z_{cs}(3985)$ \cite{BESIII:2020qkh,BESIII:2022qzr}, $Z_{cs}(4000)$ \cite{LHCb:2021uow,LHCb:2023hxg}, and $Z_{cs}(4220)$ \cite{LHCb:2021uow} listed in Table \ref{Exp} are reported with high significances. On the contrary, due to the low significance, the width of the $Z_{cs}(4123)$ state is not extracted yet \cite{BESIII:2022vxd}, this state still needs further confirmation. The observed exotic $Z_{cs}$ states also have various interpretations, including the molecular states \cite{Chen:2020yvq,Yan:2021tcp,Meng:2021rdg,Zhai:2022ied,Wu:2021cyc,Du:2022jjv,Chen:2021erj,Cheng:2023vyv}, tetraquark states \cite{Wang:2023vtx,Maiani:2021tri,Shi:2021jyr,Jin:2020yjn,Yang:2021zhe}, mixing schemes \cite{Karliner:2021qok,Han:2022fup,Cao:2022rjp}, and cusp effects \cite{Luo:2022xjx}. Since the $Z_{cs}(3985)^+$ and $Z_{cs}(4000)^+$ are both close to the threshold of $D_s^+\bar{D}^{*0}+D_s^{*+}\bar{D}^0$ system, the question of whether the $Z_{cs}(3985)$ and $Z_{cs}(4000)$ are the same state \cite{Wu:2021cyc,Ortega:2021enc,Giron:2021sla} or two different ones \cite{Wang:2023vtx,Maiani:2021tri,Meng:2021rdg,Chen:2021erj,Yang:2021zhe,Han:2022fup,Cheng:2023vyv} is still under debate. Particularly, a recent investigation from the BESIII collaboration reported the absence of the $Z_{cs}(3985)$ in the $J/\Psi K$ final states \cite{BESIII:2023wqy}. This result favor the view that the $Z_{cs}(3985)$ and $Z_{cs}(4000)$ are two different states. According to the different observed channels or the heavy quark spin symmetry \cite{Meng:2021rdg}, the $Z_{cs}(4000)$ could be assigned as the SU(3)$_{\text{f}}$ partner of the $Z_{c}(3900)$. However, such assignment leads to two difficulties:
\begin{itemize}
\item[(1)] The width of the $Z_{cs}(4000)$ is about ten times larger than that of the $Z_{c}(3900)$.
\item[(2)] The SU(3) symmetry requires the existence of a $Z_c$ that is close to the $D\bar{D}^{*}$ threshold with $J^{PC}=1^{++}$, this state should be the SU(3)$_{\text{f}}$ partner of the $Z_{cs}(3985)$. However, such state is missing in experiment.
\end{itemize}


Since the masses of these discussed $Z_c$ and $Z_{cs}$ states are all slightly above their corresponding thresholds in the $D^{(*)}\bar{D}^{(*)}$ and $D^{(*)}\bar{D}^{(*)}_s$ systems, respectively, this important feature leads the molecule resonance picture becomes a natural interpretation to these states.

In this work, we assume that the discussed $Z_c$ and $Z_{cs}$ states are resonances composed of the $D^{(*)}\bar{D}^{(*)}$ and $D_s^{(*)}\bar{D}^{(*)}$ components, respectively, and explore a unified effective field theory to describe their masses and widthes. The effective potentials of the $Z_c$ and $Z_{cs}$ states are constructed by analogy with the leading-order (LO) and next-to-leading-order (NLO) $N-\bar{N}$ effective potentials \cite{Kang:2013uia}. The involved LECs are determined with the data from the observed $Z_c$ and $Z_{cs}$ states. The effective potentials of the $Z_c$ and $Z_{cs}$ states with different quantum numbers can be related with respect to the HQS and SU(3) flavor symmetry. We will show that this framework is promising for a unified description of the $Z_c$ and $Z_{cs}$ after considering a simplified SU(3) breaking effect. In addition, this SU(3) breaking effect is also crucial to explain the the absence of a $J^{PC}=1^{++}$ $Z_c$ state and the large width difference between the $Z_c(3900)$ and $Z_{cs}(4000)$.

This paper is organised as follows. We present our theoretical
framework in Sec.~\ref{Sec 2}. In Sec.~\ref{Sec 3}, we present our two calculations to discuss the differences and similarities between the $Z_{cs}$ and $Z_c$ states, we will also present our numerical results and discussions in this section. Sec.~\ref{Sec 4} is a summary.



\section{Theoretical framework}\label{Sec 2}
Firstly, we present the wave functions of the considered $Z_{cs}$ and $Z_{c}$ systems. In each system, there are six $S$-wave states, we collectively express them as
\begin{eqnarray}
&\left|D\bar{D}_{(s)};0^{+\tilde{+}}\right\rangle&=\left|D\bar{D}_{(s)}\right\rangle_{J=0},\label{1}\\
&\left|D\bar{D}_{(s)}^*;1^{+\tilde{+}}\right\rangle&=\left(\left|D\bar{D}^*_{(s)}\right\rangle_{J=1}+\left|D^*\bar{D}_{(s)}\right\rangle_{J=1}\right)/\sqrt{2},\nonumber\\\label{2}\\
&\left|D\bar{D}_{(s)}^*;1^{+\tilde{-}}\right\rangle&=\left(\left|D\bar{D}^*_{(s)}\right\rangle_{J=1}-\left|D^*\bar{D}_{(s)}\right\rangle_{J=1}\right)/\sqrt{2},\nonumber\\\label{3}\\
&\left|D^*\bar{D}^*_{(s)};0^{+\tilde{+}}\right\rangle&=\left|D^*\bar{D}^*_{(s)}\right\rangle_{J=0},\label{4}\\
&\left|D^*\bar{D}^*_{(s)};1^{+\tilde{-}}\right\rangle&=\left|D^*\bar{D}^*_{(s)}\right\rangle_{J=1},\label{5}\\
&\left|D^*\bar{D}^*_{(s)};2^{+\tilde{+}}\right\rangle&=\left|D^*\bar{D}^*_{(s)}\right\rangle_{J=2}.\label{6}
\end{eqnarray}
Here, the superscript ``$\sim$'' on the $C$-parity number is only for the $D^{(*)}\bar{D}_s^{(*)}$ state, since the $D^{(*)}\bar{D}_s^{(*)}$ state does not have the $C$-parity, we use the $|D^{(*)}D_{s}^{(*)};J^{P\tilde{C}}\rangle$ to denote that it is the strangeness partner of the $|D^{(*)}\bar{D}^{(*)};J^{PC}\rangle$ state. The ($|D\bar{D}^*;1^{+-}\rangle$, $|D^*\bar{D}^*;1^{+-}\rangle$), ($|D\bar{D}_s^*;1^{+\tilde{-}}\rangle$, $|D^*\bar{D}_s^*;1^{+\tilde{-}}\rangle$), ($|D\bar{D}^*;1^{++}\rangle$, $|D^*\bar{D}^*;2^{++}\rangle$), and
($|D\bar{D}_s^*;1^{+\tilde{+}}\rangle$, $|D^*\bar{D}_s^*;2^{+\tilde{+}}\rangle$) are the four sets of HQS doublets.

Now we construct an effective theory to describe the interactions of the $Z_{cs}$ and $Z_{c}$ states. By analogy with the $N\bar{N}$ interaction \cite{Kang:2013uia}, we introduce the leading order contact terms to describe the exchanges of $q\bar{q}$ light mesons in the $D^{(*)}\bar{D}^{(*)}$ and $D^{(*)}\bar{D}_s^{(*)}$ systems
\begin{eqnarray}
V^{\text{LO}}_{q\bar{q}}&=&\tilde{g}_s\bm{F}_1\cdot\bm{F}_2+\tilde{g}_a\bm{F}_1\cdot\bm{F}_2\bm{\sigma}_1\cdot\bm{\sigma}_2,\label{LOqq}
\end{eqnarray}
respectively. Here, the $\bm{F}_1\cdot\bm{F}_2=-\sum_{l=1}^8\lambda_1^l\lambda_2^{*l}$ and $\bm{\sigma}_1\cdot\bm{\sigma}_2=\sum_{m=1}^3\sigma_1^m\sigma_2^m$ are the flavor and spin operators of the light quark components, respectively. The $\tilde{g}_s$ and $\tilde{g}_a$ are the two LO low energy constants (LECs). Note that
\begin{eqnarray}
\bm{\lambda}_1\cdot\bm{\lambda}^*_2=\lambda_1^8\lambda_2^{*8}+\lambda_1^{i}\lambda_2^{*i}+\lambda_1^{j}\lambda_2^{*j},
\end{eqnarray}
where $i$ and $j$ sum from 1 to 3 and 4 to 7, respectively. The matrix elements of the operators $\lambda_1^8\lambda_2^{*8}$ ($\lambda_1^8\lambda_2^{*8}\bm{\sigma}_1\cdot\bm{\sigma}_2$), $\lambda_1^i\lambda_2^i$ ($\lambda_1^i\lambda_2^{*i}\bm{\sigma}_1\cdot\bm{\sigma}_2$), and $\lambda_1^j\lambda_2^j$ ($\lambda_1^j\lambda_2^{*j}\bm{\sigma}_1\cdot\bm{\sigma}_2$) quantify the fractions of the contributions from the exchanges of the isospin singlet, triplet, and two doublets light scalar (axial-vector) meson currents, respectively.

The effective potential defined in Eq. (\ref{LOqq}) allows the exchanges of two sets of light mesons with quantum numbers $I(J^P)=0(0)^+)$, $1(0^+)$, $1/2(0^+)$ and $I(J^P)=0(1^+)$, $1(1^+)$, $1/2(1^+)$. For each exchanged meson current, their spin and flavor structures are identified by the corresponding spin and flavor matrix elements, respectively. Then we use the coupling constants $\tilde{g}_s\approx g_s^2/m^2_{s}$ and $\tilde{g}_a\approx g_a^2/m^2_{a}$ to collectively quantify the dynamical effects from the exchange of each scalar and axial-vector meson current \cite{Chen:2022wkh}. In the SU(3) limit, the couplings $\tilde{g}_s$ ($\tilde{g}_a$) for the exchanges of scalar (axial-vector) meson currents with different isospins are the same.

Then we proceed to introduce the effective potentials of the $Z_{cs}$ and $Z_{c}$ states at NLO. By analogy with the NLO $N-\bar{N}$ interaction \cite{Kang:2013uia}, the contact terms that attributed to the exchanges of light mesons in the $Z_{cs}$ and $Z_{c}$ systems read
\begin{eqnarray}
V^{\text{NLO}}_{q\bar{q}}&=&\left(\bm{F}_1\cdot\bm{F}_2\right)\big[g_1\bm{q}^2+g_2\bm{k}^2+\left(g_3\bm{q}^2+g_4\bm{k}^2\right)\bm{\sigma}_1\cdot\bm{\sigma}_2\nonumber\\
&&+\frac{i}{2}g_5\left(\bm{\sigma}_1+\bm{\sigma}_2\right)\cdot\left(\bm{q}\times\bm{k}\right)
+g_6\left(\bm{q}\cdot\bm{\sigma}_1\right)\left(\bm{q}\cdot\bm{\sigma}_2\right)\nonumber\\&&+g_7\left(\bm{k}\cdot\bm{\sigma}_1\right)\left(\bm{k}\cdot\bm{\sigma}_2\right)\big].\label{VNLO}
\end{eqnarray}
Here, the transferred momentum $\bm{q}=\bm{p}-\bm{p}^\prime$, and the average momentum $\bm{k}$ is defined by $\bm{k}=(\bm{p}^\prime+\bm{p})/2$. When performing an $S$-wave channel partial-wave projection, the $g_5$ term will not contribute to the $S$-wave potential, and the possible $\bm{p}\cdot\bm{p}^\prime$ terms will also vanish, then Eq. (\ref{VNLO}) can be rewritten as
\begin{eqnarray}
V^{\text{NLO}}_{q\bar{q}}&=&\left(\bm{F}_1\cdot\bm{F}_2\right)\left(\bm{p}^2+\bm{p}^{\prime2}\right)\big[\left(\tilde{g}_1+\tilde{g}_2\right)\nonumber\\&&+\left(\tilde{g}_3+\tilde{g}_4+\tilde{g}_6+\tilde{g}_7\right)\left(\bm{\sigma}_1\cdot\bm{\sigma}_2\right)\big]\nonumber\\
&=&\left(\bm{F}_1\cdot\bm{F}_2\right)[\tilde{g}_{sp}+\tilde{g}_{ap}\left(\bm{\sigma}_1\cdot\bm{\sigma}_2\right)]\left(\bm{p}^2+\bm{p}^{\prime2}\right).\label{NLOqq}
\end{eqnarray}
The LECs $\tilde{g}_{sp}\approx g_{sp}^{2}/m^{2}_{sp}$ and $\tilde{g}_{ap}\approx g_{ap}^{2}/m^{2}_{ap}$ are introduced to describe the effective couplings of the momentum dependent terms $\bm{F}_1\cdot\bm{F}_2(\bm{p}^2+\bm{p}^{\prime2})$ and $\bm{F}_1\cdot\bm{F}_2(\bm{\sigma}_1\cdot\bm{\sigma}_2)(\bm{p}^2+\bm{p}^{\prime2})$, respectively.


By collecting Eq. (\ref{LOqq}) with Eq. (\ref{NLOqq}), the total effective potential of the $Z_{cs}$ or $Z_{c}$ states can be written as
\begin{eqnarray}
V_{Z_{c(s)}}&=&V_{q\bar{q}}^{\text{LO}}+V_{q\bar{q}}^{\text{NLO}}.\label{Vtotal}
\end{eqnarray}
This effective potential includes four LECs, i.e., the $\tilde{g}_s$, $\tilde{g}_a$, $\tilde{g}_{sp}$, and $\tilde{g}_{ap}$, we will determine them in the next section.

\section{Numerical results and discussions}\label{Sec 3}
In this section, we firstly introduce our scheme to include the SU(3) breaking effect in the $Z_{cs}$ and $Z_{c}$ states, then we will perform two calculations to compare the differences and similarities of the observed $Z_{cs}$ and $Z_c$ states in Sec. \ref{Sec 3B} and Sec. \ref{Sec 3C}, respectively.
\subsection{SU(3) breaking factor $g_x$}\label{Sec 3A}
In Table \ref{Operators}, for the considered $Z_{cs}$ and $Z_c$ states, we list their matrix elements of the operators
\begin{eqnarray}
\mathcal{O}_1=\sum_{i=1}^7\lambda_1^i\cdot\lambda_2^i,\quad&&
\mathcal{O}_2=\lambda^8_1\cdot \lambda^8_2,\nonumber\\
\mathcal{O}_3=\sum_{i=1}^7\lambda_1^i\cdot\lambda_2^i\bm{\sigma}_1\cdot\bm{\sigma}_2,\quad&&
\mathcal{O}_4=\lambda_1^8\lambda_2^8\bm{\sigma}_1\cdot\bm{\sigma}_2.\nonumber
\end{eqnarray}
We can write out the effective potential of a specific $Z_{cs}$ or $Z_{c}$ state from Table \ref{Operators} and Eq. (\ref{Vtotal}).

\begin{table}[htbp]
\renewcommand\arraystretch{1.5}
\caption{The matrix elements of the operators $\mathcal{O}_1$ ($\sum_{i=1}^7\lambda_1^i\cdot\lambda_2^i$)
$\mathcal{O}_2$ ($\lambda^8_1\cdot \lambda^8_2$), $\mathcal{O}_3$ ($\sum_{i=1}^7\lambda_1^i\cdot\lambda_2^i\bm{\sigma}_1\cdot\bm{\sigma}_2$), $\mathcal{O}_4$ ($\lambda_1^8\lambda_2^8\bm{\sigma}_1\cdot\bm{\sigma}_2$) in the $Z_c$ ($D^{(*)}\bar{D}^{(*)}$) and $Z_{cs}$ ($D^{(*)}\bar{D}_{s}^{(*)}$) states. Here, we use the notation $|D^{(*)}\bar{D}_s^{(*)};J^{P\tilde{C}}\rangle$ to denote that it is the strangeness partner of the $|D^{(*)}\bar{D}^{(*)};J^{PC}\rangle$ state.\label{Operators}}
\setlength{\tabcolsep}{0.85mm}{
\begin{tabular}{ccccccccccccccc}
\toprule[1pt]
State&$\mathcal{O}_1$&$\mathcal{O}_2$&$\mathcal{O}_3$&$\mathcal{O}_4$&State&$\mathcal{O}_1$&$\mathcal{O}_2$&$\mathcal{O}_3$&$\mathcal{O}_4$\\
\hline
\hline
$|D\bar{D};0^{++}\rangle$&1&$-\frac{1}{3}$&0&0&
$|D\bar{D}_{s};0^{+\tilde{+}}\rangle$&0&$\frac{2}{3}$&0&0\\
$|D\bar{D}^*;1^{++}\rangle$&1&$-\frac{1}{3}$&1&$-\frac{1}{3}$&
$|D\bar{D}^*_{s};1^{+\tilde{+}}\rangle$&0&$\frac{2}{3}$&0&$\frac{2}{3}$\\
$|D\bar{D}^*;1^{+-}\rangle$&1&$-\frac{1}{3}$&$-1$&$\frac{1}{3}$&
$|D\bar{D}^*_{s};1^{+\tilde{-}}\rangle$&0&$\frac{2}{3}$&$0$&$-\frac{2}{3}$\\
$|D^*\bar{D}^*;0^{++}\rangle$&1&$-\frac{1}{3}$&$-2$&$\frac{2}{3}$&
$|D^*\bar{D}^*_{s};0^{+\tilde{+}}\rangle$&0&$\frac{2}{3}$&$0$&$-\frac{4}{3}$\\
$|D^*\bar{D}^*;1^{+-}\rangle$&1&$-\frac{1}{3}$&$-1$&$\frac{1}{3}$&
$|D^*\bar{D}^*_{s};1^{+\tilde{-}}\rangle$&0&$\frac{2}{3}$&$0$&$-\frac{2}{3}$\\
$|D^*\bar{D}^*;2^{++}\rangle$&1&$-\frac{1}{3}$&1&$-\frac{1}{3}$&
$|D^*\bar{D}^*_{s};2^{+\tilde{+}}\rangle$&0&$\frac{2}{3}$&0&$\frac{2}{3}$\\
\hline
\bottomrule[1pt]
\end{tabular}}
\end{table}

As presented in Table \ref{Operators}, on the one hand, in the SU(3) limit, the $Z_c$ and $Z_{cs}$ states with the same quantum numbers share identical matrix elements $\langle\mathcal{O}_1\rangle+\langle\mathcal{O}_2\rangle$ and $\langle\mathcal{O}_3\rangle+\langle\mathcal{O}_4\rangle$. Correspondingly, the $Z_c$ and $Z_{cs}$ states with the same $J^{P\tilde{C}}$ share identical effective potential, this is the requirement from the SU(3) flavor symmetry.

On the other hand, in the $Z_{cs}$ states, the matrix elements $\mathcal{O}_1$ and $\mathcal{O}_3$ vanish, the total effective potentials of the $Z_{cs}$ states consist of the operators $\mathcal{O}_2$ ($\lambda_1^8\cdot\lambda_2^8$) and $\mathcal{O}_4$ ($\lambda_1^8\cdot\lambda_2^8\bm{\sigma}_1\cdot\bm{\sigma}_2$). They correspond to the interactions from the exchanges of light mesons with $n\bar{n}+s\bar{s}$ ($n=u$, $d$) components. In the $Z_c$ states, their effective potentials have non-zero contributions from the operators $\mathcal{O}_1$ ($\sum_{i=1}^7\lambda_1^i\cdot\lambda_2^i$) and $\mathcal{O}_3$ ($\sum_{i=1}^7\lambda_1^i\cdot\lambda_2^i\bm{\sigma}_1\cdot\bm{\sigma}_2$) (except the $\langle \mathcal{O}_3 \rangle=0$ in the $|D\bar{D};0^{++}\rangle$ state). Specifically, the matrix elements in the operators $\mathcal{O}_1$ and $\mathcal{O}_3$ with $i=1,2,3$ are non-zero, they correspond to the interactions from the exchanges of $n\bar{n}$ ($n=u$, $d$) light mesons, while the matrix elements in the operators $\mathcal{O}_1$ and $\mathcal{O}_3$ with $i=4,5,6,7$ are zero, they correspond to the interactions from the exchanges of $n\bar{s}$/$s\bar{n}$ strange mesons.

To include the SU(3) breaking effects between the $Z_{cs}$ and $Z_{c}$ states, from the effective potentials listed in Table \ref{Operators}, we adopt the following approximation
\begin{eqnarray}
m_{i}<m_{8},
\end{eqnarray}
Here, the $m_{i}$ and $m_{8}$ denote the masses of exchanged light mesons that are related to the flavor isospin triplet $\sum_{i=1}^{3}\lambda^i_1\cdot\lambda_2^i$ and isospin singlet $\lambda_1^8\cdot\lambda_2^8$ operators, respectively. Comparing to the effective coupling constants $\tilde{g}_s$, $\tilde{g}_a$, $\tilde{g}_{sp}$, $\tilde{g}_{ap}$ that are related to the $\lambda_1^8\cdot\lambda_2^8$ and proportional to $1/m_{8}^2$, the effective coupling constants $\tilde{g}^\prime_s$, $\tilde{g}^\prime_a$, $\tilde{g}^\prime_{sp}$, $\tilde{g}^\prime_{ap}$ that are related to the $\lambda_1^i\cdot\lambda_2^i$ operator are magnified by $1/m_{i}^2$, i.e.,
\begin{eqnarray}
\tilde{g}_s=\frac{g_s^2}{m_{s8}^2}<\tilde{g}_s^\prime=\frac{g_s^2}{m^2_{si}},\label{grelation1}\\
\tilde{g}_a=\frac{g_a^2}{m_{a8}^2}<\tilde{g}_a^\prime=\frac{g_a^2}{m^2_{ai}},\label{grelation2}\\
\tilde{g}_{sp}=\frac{g_{sp}^2}{m_{sp8}^2}<\tilde{g}_{sp}^\prime=\frac{g_{sp}^2}{m^2_{spi}},\label{grelation3}\\
\tilde{g}_{ap}=\frac{g_{ap}^2}{m_{ap8}^2}<\tilde{g}_{ap}^\prime=\frac{g_{ap}^2}{m^2_{api}}.\label{grelation4}
\end{eqnarray}
In principle, the SU(3) breaking effects should be slightly different among the four interacting terms that are related to the $\tilde{g}_{s}/\tilde{g}_{s}^\prime$, $\tilde{g}_{a}/\tilde{g}_{a}^\prime$, $\tilde{g}_{sp}/\tilde{g}_{sp}^\prime$, $\tilde{g}_{ap}/\tilde{g}_{ap}^\prime$ couplings. At present, we do not have enough data to specify such differences. Instead, we introduce a global SU(3) breaking factor $g_x$, and redefine the $\tilde{g}_{s}^\prime$, $\tilde{g}_{a}^\prime$, $\tilde{g}_{sp}^\prime$, $\tilde{g}_{ap}^\prime$ as
\begin{eqnarray}
\tilde{g}^\prime_{s}=g_x\tilde{g}_{s},\,\,
\tilde{g}^\prime_{a}=g_x\tilde{g}_{a},\,\,
\tilde{g}^\prime_{sp}=g_x\tilde{g}_{sp},\,\,
\tilde{g}^\prime_{ap}=g_x\tilde{g}_{ap}.\label{gxrelation}
\end{eqnarray}
According to Eqs. (\ref{grelation1}-\ref{grelation4}), the factor $g_x$ is expected to have $g_x>1$.
\subsection{The differences of the $Z_{cs}$ and $Z_c$ states}\label{Sec 3B}
After identifying the SU(3) breaking effect, we present our first calculation to clarify the differences between the $Z_{cs}$ and $Z_c$ states. Here, we need to pin down the five coupling parameters, i.e., the $\tilde{g}_s$, $\tilde{g}_a$, $\tilde{g}_{sp}$, $\tilde{g}_{ap}$, and $g_x$, while the $\tilde{g}^\prime_s$, $\tilde{g}^\prime_a$, $\tilde{g}^\prime_{sp}$, $\tilde{g}^\prime_{ap}$ can be obtained from the above five parameters as redefined in Eqs. (\ref{gxrelation}).

The $Z_{cs}(3985)$ \cite{BESIII:2020qkh,BESIII:2022qzr} and $Z_{cs}(4000)$ \cite{LHCb:2021uow,LHCb:2023hxg} are reported in the $D_s^+\bar{D}^{*0}+D_s^{*+}\bar{D}^0$ ($D_s^+D^{*-}+D_s^{*+}D^{-}$) and $J/\Psi K^+$ ($J/\Psi K_S^0$) final states, respectively. 
In this work, we treat the $Z_{cs}(4000)$ and $Z_{cs}(3985)$ as two different states.
According to the heavy quark spin symmetry \cite{Meng:2021rdg}, the $|D\bar{D}_s^*;1^{+\tilde{+}}\rangle$ state can not decay into the $J/\Psi K^+$ final states. Thus, we assume the following assignments
\begin{eqnarray}
\text{Set}\,1: Z_{cs}(4000) |D\bar{D}_s^*;1^{+\tilde{-}}\rangle,\,&&Z_{cs}(3985)
|D\bar{D}_s^*;1^{+\tilde{+}}\rangle.\nonumber
\end{eqnarray}
We denote this set of assignments as Set 1. From Table \ref{Operators}, we can directly obtain the effective potentials of the $Z_{cs}(4000)$ and $Z_{cs}(3985)$ as
\begin{eqnarray}
V_{Z_{cs}(4000)}&=&\frac{2}{3}(\tilde{g}_s-\tilde{g}_a)+\frac{2}{3}(\tilde{g}_{sp}-\tilde{g}_{ap})\left(p^2+p^{\prime2}\right),\label{Zcs4000}\\
V_{Z_{cs}(3985)}&=&\frac{2}{3}(\tilde{g}_s+\tilde{g}_a)+\frac{2}{3}(\tilde{g}_{sp}+\tilde{g}_{ap})\left(p^2+p^{\prime2}\right).\label{Zcs3985}
\end{eqnarray}
As presented in Table \ref{Operators}, the matrix elements $\langle\mathcal{O}_1\rangle$ and $\langle\mathcal{O}_3\rangle$ vanish in the $Z_{cs}$ states, thus, we do not need to introduce the $g_x$ to describe the effective potentials of the $Z_{cs}$ states with different quantum numbers. 

By introducing the experimental masses and widthes of the $Z_{cs}(4000)$ and $Z_{cs}(3985)$, we can solve the LECs ($\tilde{g}_s$, $\tilde{g}_a$, $\tilde{g}_{sp}$, $\tilde{g}_{ap}$) with the following Lippmann-Schwinger equation
\begin{eqnarray}
T\left(\bm{p}^\prime,\bm{p}\right)&=&V\left(\bm{p}^\prime,\bm{p}\right)\nonumber\\&&+\int\frac{d^3\bm{q}}{(2\pi)^3}\frac{V(\bm{p}^\prime,\bm{q})T(\bm{q},\bm{p})u^2\left(\Lambda\right)}{E-\sqrt{m_{1}^2+\bm{q}^2}+\sqrt{m_2^2+\bm{q}^2}},\nonumber\\
\end{eqnarray}
where $m_1$ and $m_2$ are the masses of the charmed and charmed-strange meson components in the $Z_{cs}$ states. To suppress the contributions from high momenta, we introduce a dipole form factor $u(\Lambda)=(1+q^2/\Lambda^2)^{-2}$, and set $\Lambda=1.0$ GeV \cite{Nakamura:2022jpd,Leinweber:2003dg,Wang:2007iw,Chen:2022wkh}, we will discuss the $\Lambda$-dependences of our results in Sec. \ref{Sec 3C}.

For the separable effective potentials in Eqs. (\ref{Zcs4000}-\ref{Zcs3985}), we solve the above LSE with the matrix-inversion method \cite{Epelbaum:2017byx}. The conditions that the poles of the $|D\bar{D}_{s}^*;1^{+\tilde{-}}\rangle$ and $|D\bar{D}_s^*;1^{+\tilde{-}}\rangle$ states can coexist are
\begin{eqnarray}
\text{Det}\left[\left(
  \begin{array}{cc}
    1 & 0 \\
    0 & 1 \\
  \end{array}
\right)-\left(
  \begin{array}{cc}
    A & B \\
    B & 0 \\
  \end{array}
\right)\left(
  \begin{array}{cc}
    G_0 & G_2 \\
    G_2 & G_4 \\
  \end{array}
\right)\right]=0,\label{1pmzero}\\
\text{Det}\left[\left(
  \begin{array}{cc}
    1 & 0 \\
    0 & 1 \\
  \end{array}
\right)-\left(
  \begin{array}{cc}
    C & D \\
    D & 0 \\
  \end{array}
\right)\left(
  \begin{array}{cc}
    G_0 & G_2 \\
    G_2 & G_4 \\
  \end{array}
\right)\right]=0,\label{1ppzero}
\end{eqnarray}
with $A=\frac{2}{3}(\tilde{g}_s-\tilde{g}_a)$, $B=\frac{2}{3}(\tilde{g}_{sp}-\tilde{g}_{ap})$, $C=\frac{2}{3}(\tilde{g}_s+\tilde{g}_a)$, and $D=\frac{2}{3}(\tilde{g}_{sp}+\tilde{g}_{ap})$.
Here, $G_n$ is defined as
\begin{eqnarray}
G_n=\int \frac{1}{2\pi^2}\frac{q^{2+n}(1+\frac{q^2}{\Lambda})^{-4}}{E-\sqrt{m_1^2+q^2}-\sqrt{m_2^2+q^2}}.
\end{eqnarray}
We replace the integration variable $q$ with $q\rightarrow q\times \text{exp}(-i\theta)$, and set $0<\theta<\frac{\pi}{2}$ to search for the $|D\bar{D}_s^*;1^{+-}\rangle$ and $|D^*\bar{D}_s^*;1^{++}\rangle$ resonances in the second Riemann sheet. By adopting the experimental central values of the $Z_{cs}(4000)$ \cite{LHCb:2021uow} and $Z_{cs}(3985)$ \cite{BESIII:2020qkh}, the four LECs are obtained as
\begin{eqnarray}
\tilde{g}_s&=&32.1 \,\,\text{GeV}^{-2}, \tilde{g}_a=20.8 \,\,\text{GeV}^{-2},\\
\tilde{g}_{sp}&=&-61.5 \,\,\text{GeV}^{-4}, \tilde{g}_{ap}=-22.8 \,\,\text{GeV}^{-4}.
\end{eqnarray}

As presented in Eq. (\ref{gxrelation}), in the $Z_c$ system, we introduce a factor $g_x$ to redefine the LECs ($\tilde{g}_s$, $\tilde{g}_a$, $\tilde{g}_{sp}$, $\tilde{g}_{ap}$) that are related to the operators $\mathcal{O}_2$ and $\mathcal{O}_4$, this factor is only related to the effective potentials of $Z_c$ states. Explicitly, the effective potentials of $|D\bar{D}^*;1^{+-}\rangle$ ($Z_c(3900)$) and $|D\bar{D}^*;1^{++}\rangle$ states can be written as
\begin{eqnarray}
V_{|D\bar{D}^*;1^{+-}\rangle}&=&\left(g_x-\frac{1}{3}\right)\left(\tilde{g}_s-\tilde{g}_a\right)\nonumber\\&&+\left(g_x-\frac{1}{3}\right)
\left(\tilde{g}_{sp}-\tilde{g}_{ap}\right)\left(p^2+p^{\prime2}\right),\\
V_{|D\bar{D}^*;1^{++}\rangle}&=&\left(g_x-\frac{1}{3}\right)\left(\tilde{g}_s+\tilde{g}_a\right)\nonumber\\&&+\left(g_x-\frac{1}{3}\right)
\left(\tilde{g}_{sp}+\tilde{g}_{ap}\right)\left(p^2+p^{\prime2}\right).
\end{eqnarray}
In the HQS limit, the $|D^*\bar{D}^*;1^{+-}\rangle$ (we assume this state corresponds to the $Z_c(4020)$) and $|D^{*}\bar{D}^*;2^{++}\rangle$ share identical effective potentials to that of the $|D\bar{D}^*;1^{+-}\rangle$ and $|D\bar{D}^*;1^{++}\rangle$ states, respectively. Thus, We no longer list the effective potentials of the $|D^*\bar{D}^*;1^{+-}\rangle$ and $|D^{*}\bar{D}^*;2^{++}\rangle$ states further.

To pin down the SU(3) breaking factor $g_x$, we adopt the obtained four LECs extracted from the $Z_{cs}(4000)$ and $Z_{cs}(3985)$ states to the effective potentials of the ($|D\bar{D}^*;1^{+-}\rangle$, $|D^*\bar{D}^*;1^{+-}\rangle$) and ($|D\bar{D}^*;1^{++}\rangle$, $|D^*\bar{D}^*;2^{++}\rangle$) states, we run the $g_x$ in a reasonable region, then we check the behaviors of the masses and widthes of these two sets of HQS doublets. We find the possible $g_x$ region by reproducing the experimental resonance parameters \cite{ParticleDataGroup:2022pth} of the $Z_c(3900)$ (or the $Z_c(4020)$).

\begin{figure*}[!htbp]
    \centering
    \includegraphics[width=0.9\linewidth]{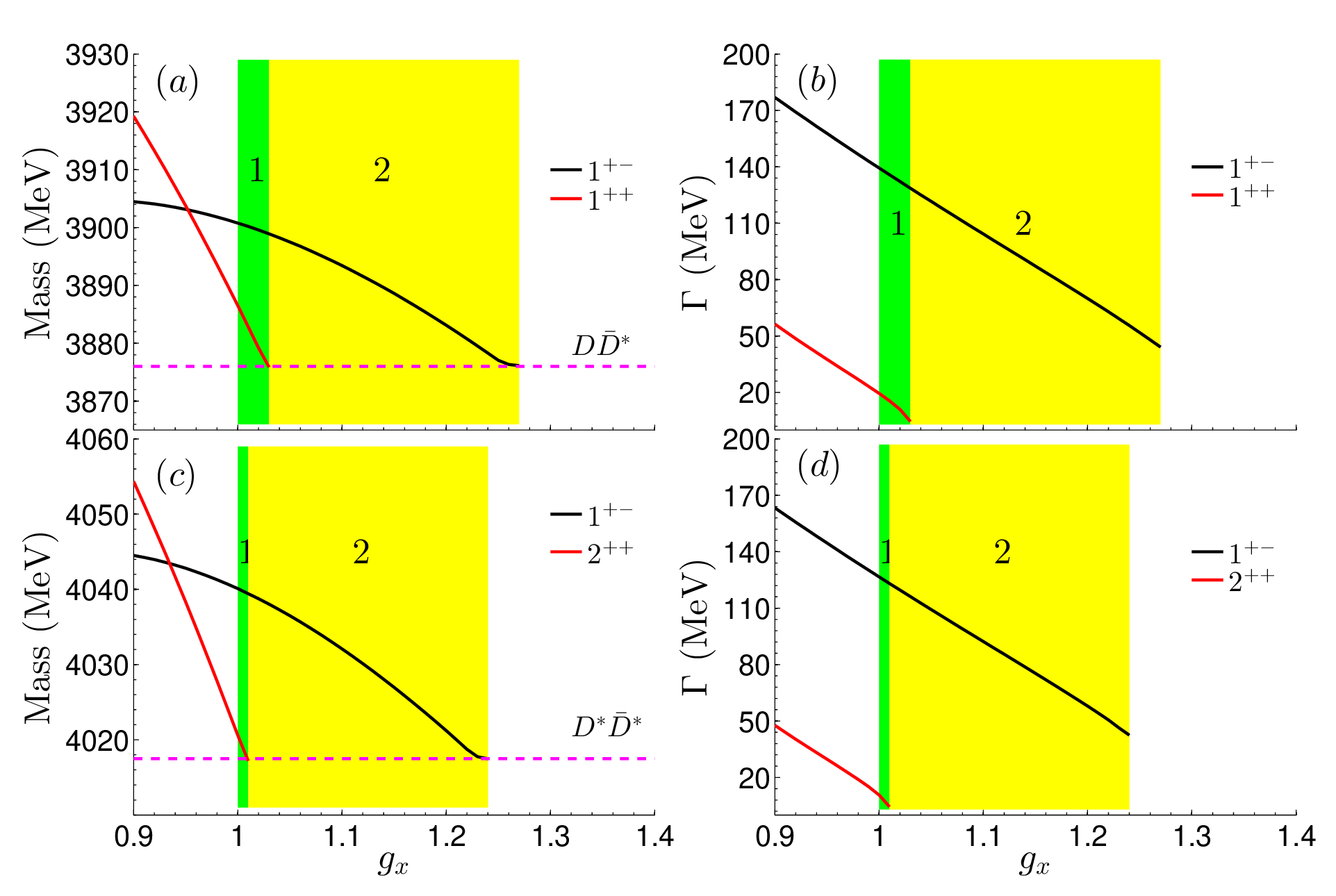}
    \caption{The $g_x$-dependences of the masses and widthes of the ($|D\bar{D}^*;1^{+-}\rangle$, $|D^*\bar{D}^*;1^{+-}\rangle$) (black lines) and ($|D\bar{D}^*;1^{++}\rangle$, $|D^*\bar{D}^*;2^{++}\rangle$) (red lines) doublets. The LECs are extracted from the experimental inputs in Set 1.}
    \label{ZcSU3}
\end{figure*}

The $g_x$-dependences of the masses and widthes of the ($|D\bar{D}^*;1^{+-}\rangle$, $|D^*\bar{D}^*;1^{+-}\rangle$) and ($|D\bar{D}^*;1^{++}\rangle$, $|D^*\bar{D}^*;2^{++}\rangle$) doublets are presented in Fig. \ref{ZcSU3}. As discussed in Sec. \ref{Sec 3A}, the SU(3) breaking factor $g_x$ is expected to be greater than 1, here, in order to show the evolutions of the masses and widthes of these two sets of HQS doublets with $g_x$, we run $g_x$ from 0.9, slightly smaller than its lower limit.

We firstly discuss the results in the SU(3) limit at $g_x=1.0$. As listed in Table \ref{SU3limit}, the obtained resonance parameters of the $Z_c$ states with $J^{PC}=1^{+-}$ and $1^{++}$ are very similar to that of the observed $Z_{cs}(4000)$ and $Z_{cs}(3985)$ states, respectively.
\begin{table}[htbp]
\renewcommand\arraystretch{1.5}
\caption{The comparison of the masses and widthes between the $Z_c$ and $Z_{cs}$ states with $J^{P\tilde{C}}=1^{+\tilde{\pm}}$. We adopt the LECs solved from the inputs in Set 1 to calculate the $Z_c$ states. The SU(3) breaking factor $g_x$ is fixed at 1.0 in the SU(3) limit.\label{SU3limit}}
\setlength{\tabcolsep}{0.85mm}{
\begin{tabular}{ccccccccccccccc}
\toprule[1pt]
State&Mass (MeV)&$\delta_{M-M_\text{Tr}}$ (MeV)&Width (MeV)\\
\hline
$|DD^*;1^{++}\rangle$&3886.5&10.7&19.4\\
$|DD_s^{*};1^{+\tilde{+}}\rangle$&3985.2&5.9&$13.8$\\
$|DD^*;1^{+-}\rangle$&3900.1&24.3&139.3\\
$|DD_s^{*};1^{+\tilde{-}}\rangle$&4003.0&23.7&131.0\\
\hline
\bottomrule[1pt]
\end{tabular}}
\end{table}
Here, $\delta_{M-M_{\text{Tr}}}$ is defined as $\delta_{M-M_{\text{Tr}}}=M-M_{\text{Tr}}$, $M$ is the mass of considered $Z_c$/$Z_{cs}$ state, and $M_{\text{Tr}}$ is the corresponding two-meson threshold. The similarities of $\delta_{M-M_{\text{Tr}}}$ and widthes between the $|DD^*;1^{++}\rangle$ ($|DD^*;1^{+-}\rangle$) and $|DD_s^{*};1^{+\tilde{+}}\rangle$ ($|DD_s^{*};1^{+\tilde{-}}\rangle$) states are exactly the requirements from the SU(3) flavor symmetry.

Then we divide two $g_x$ regions, i.e., the region 1 (green band) and region 2 (yellow band) in Fig. \ref{ZcSU3} to discuss our results. The smaller and bigger $g_x$ values in region 1 and region 2 denote the tiny and considerable SU(3) breaking effects, respectively. The $g_x$-dependences of the masses and widthes of the ($|D\bar{D}^*;1^{+-}\rangle$, $|D\bar{D}^*;1^{++}\rangle$) states are presented in Fig. \ref{ZcSU3} (a) and (b), respectively. As shown in Fig. \ref{ZcSU3} (a), in region 1, where only a tiny SU(3) breaking effect is introduced, a lighter and narrower $|D\bar{D}^*; 1^{++}\rangle$ and a heavier and broader $|D\bar{D}^*;1^{+-}\rangle$ states can coexist. Besides, the masses of the $|DD^*;1^{++}\rangle$ and $|DD^*;1^{+-}\rangle$ have different $g_x$-dependent behaviors. As the $g_x$ increase, the mass of the $|D\bar{D}^{*};1^{+-}\rangle$ state decreases slowly and can cross the region 1, while the mass of the $|D\bar{D}^{*};1^{++}\rangle$ state decreases and moves to the threshold of $D\bar{D}^*$ rapidly. As the mass of $|D\bar{D}^{*};1^{++}\rangle$ state is equal to the threshold of $D\bar{D}^*$ (the $g_x$ is at the upper limit of region 1), we can no longer find the $|D\bar{D}^{*};1^{++}\rangle$ state in the second Riemann sheet.

At $g_x>1.03$, we move to the region 2, where only the $|D\bar{D}^*;1^{+-}\rangle$ state can be found in the second Riemann sheet. We find that the width of the $|D\bar{D}^*;1^{+-}\rangle$ decreases from 128 MeV to 48 MeV as the factor $g_x$ increases from 1.03 to 1.26, respectively. This result shows that a considerable SU(3) breaking can lead the width of the $|D\bar{D}^*;1^{+-}\rangle$ state become much smaller.

The above discussions show that our framework provide possible explanations to the absence of the $|D\bar{D}^*;1^{++}\rangle$ state and the large width difference between the $Z_c(3900)$ and $Z_{cs}(4000)$ states. These two questions are not expected from the SU(3) flavor symmetry but can be solved simultaneously by introducing a considerable SU(3) breaking effect. The $|D^*\bar{D}^*;1^{+-}\rangle$ and $|D^*\bar{D}^*;2^{++}\rangle$ are the HQS partners of the $|D\bar{D}^*;1^{+-}\rangle$ and $|D\bar{D}^*;1^{++}\rangle$ states, respectively. Their masses and widthes have very similar $g_x$-dependences to that of the $|D\bar{D}^*;1^{+-}\rangle$ and $|D\bar{D}^*;1^{++}\rangle$ states, respectively. We illustrate them in Fig. \ref{ZcSU3} (c-d).

Although the obtained width of the $Z_{c}(3900)$ in region 2 is still larger than the PDG \cite{ParticleDataGroup:2022pth} average value $28.4\pm2.6$ MeV, we need to emphasis that in this calculation, we use the central values of the resonance parameters of the $Z_{cs}(4000)$ \cite{LHCb:2021uow} and $Z_{cs}(3985)$ \cite{BESIII:2020qkh}, these inputs still have considerable experimental uncertainties, further measurements on the resonance parameters of the $Z_{cs}(3985)$ and $Z_{cs}(4000)$ from other experiments or processes will provide important guidances to our model.

\subsection{The similarities of the $Z_{cs}$ and $Z_c$ states}\label{Sec 3C}
Then we proceed to investigate the similarities of the $Z_{cs}$ and $Z_c$ states. 
In this subsection, within the same framework, we use another scheme to compare the $Z_c$ and $Z_{cs}$ states. 
We separately determine the parameters $\tilde{g}_s$, $\tilde{g}_a$, $\tilde{g}_{sp}$, and $\tilde{g}_{ap}$ from the experimental data of $Z_{cs}$ and $Z_{c}$ states, we label the selected $Z_{cs}$ and $Z_c$ states as Set 1 and Set 2, respectively. Then we compare the similarities of these two sets of parameters. If our framework can indeed describe the observed $Z_{cs}$ and $Z_c$ states, the LECs extracted from the $Z_{cs}$ data should be very similar to that of the $Z_c$ data. To identify their similarities, we further define a quantity $\chi$ with
\begin{eqnarray}
\chi=\sqrt{\sum_{i=1}^4(g_i^c-g_i^{cs})^2}.
\end{eqnarray}
Here, $g^{c(s)}_1=g_{s}^{c(s)}$, $g^{c(s)}_2=g_{a}^{c(s)}$, $g^{c(s)}_3=g_{sp}^{c(s)}$, and $g^{c(s)}_4=g_{ap}^{c(s)}$. We use the superscript ``$cs$" and ``$c$" to denote the parameters that are extracted from the experimental $Z_{cs}$ and $Z_c$ data, respectively.

We still select the $Z_{cs}(4000)$ and $Z_{cs}(3985)$ states to pin down the $\tilde{g}_s^{cs}$, $\tilde{g}_a^{cs}$, $\tilde{g}_{sp}^{cs}$, and $\tilde{g}_{ap}^{cs}$. To determine the $\tilde{g}_s^{c}$, $\tilde{g}_a^{c}$, $\tilde{g}_{sp}^{c}$, and $\tilde{g}_{ap}^{c}$ in the $Z_c$ sector, we need to select two observed $Z_c$ states. Here, since the $Z_c(4020)$ is the HQS partner of the $Z_c(3900)$, if we use the experimental mass and width of $Z_c(3900)$ as inputs, then the resonance parameters of $Z_c(4020)$ can no longer be regarded as independent inputs.

Alternatively, we notice that the BELLE collaboration \cite{Belle:2008qeq} reported a $Z_c(4050)$ state in the $\pi\chi_{c1}(1P)$ final states. Theoretically, this state has been discussed within the tetraquark \cite{Ebert:2008kb,Patel:2014vua,Deng:2015lca,Wang:2013llv}, hadron-molecule \cite{Liu:2009qhy,Liu:2008mi,Ding:2008gr,Lee:2008tz}, and triangle singularity \cite{Nakamura:2019emd} picture. For a more complete summary, see reviews \cite{Chen:2016qju,Albuquerque:2018jkn}. The $Z_c(4050)$ is close to the $D^*\bar{D}^*$ threshold, due to the non-observation of the $|DD^*,1^{++}\rangle$ state, its HQS partner $|D^*D^*,2^{++}\rangle$ should not exist either. Thus, we assume the $Z_c(4050)$ state is a resonance composed of the $D^*\bar{D}^*$ component with quantum number $J^{PC}=0^{++}$. Then the selected observed $Z_{cs}$ and $Z_{c}$ states in Set 1 and Set 2 are
\begin{eqnarray}
\text{Set}\,1: Z_{cs}(4000) |D\bar{D}_s^*;1^{+\tilde{-}}\rangle,\,&&Z_{cs}(3985)
|D\bar{D}_s^*;1^{+\tilde{+}}\rangle,\nonumber\\
\text{Set}\,2: Z_{c}(3900) |D\bar{D}^*;1^{+-}\rangle,\,&&Z_{c}(4050)
|D^*\bar{D}^*;0^{++}\rangle,\nonumber
\end{eqnarray}
respectively. The SU(3) breaking factor $g_x$ is related to the effective potentials of the $Z_c$ states and has not been determined yet, we fix the $\Lambda$ at 1.0 GeV, then we run the $g_x$ in a reasonable region, at the minimum $\chi$, we obtain $g_x=1.37$.

\begin{figure*}[!htbp]
    \centering
    \includegraphics[width=0.9\linewidth]{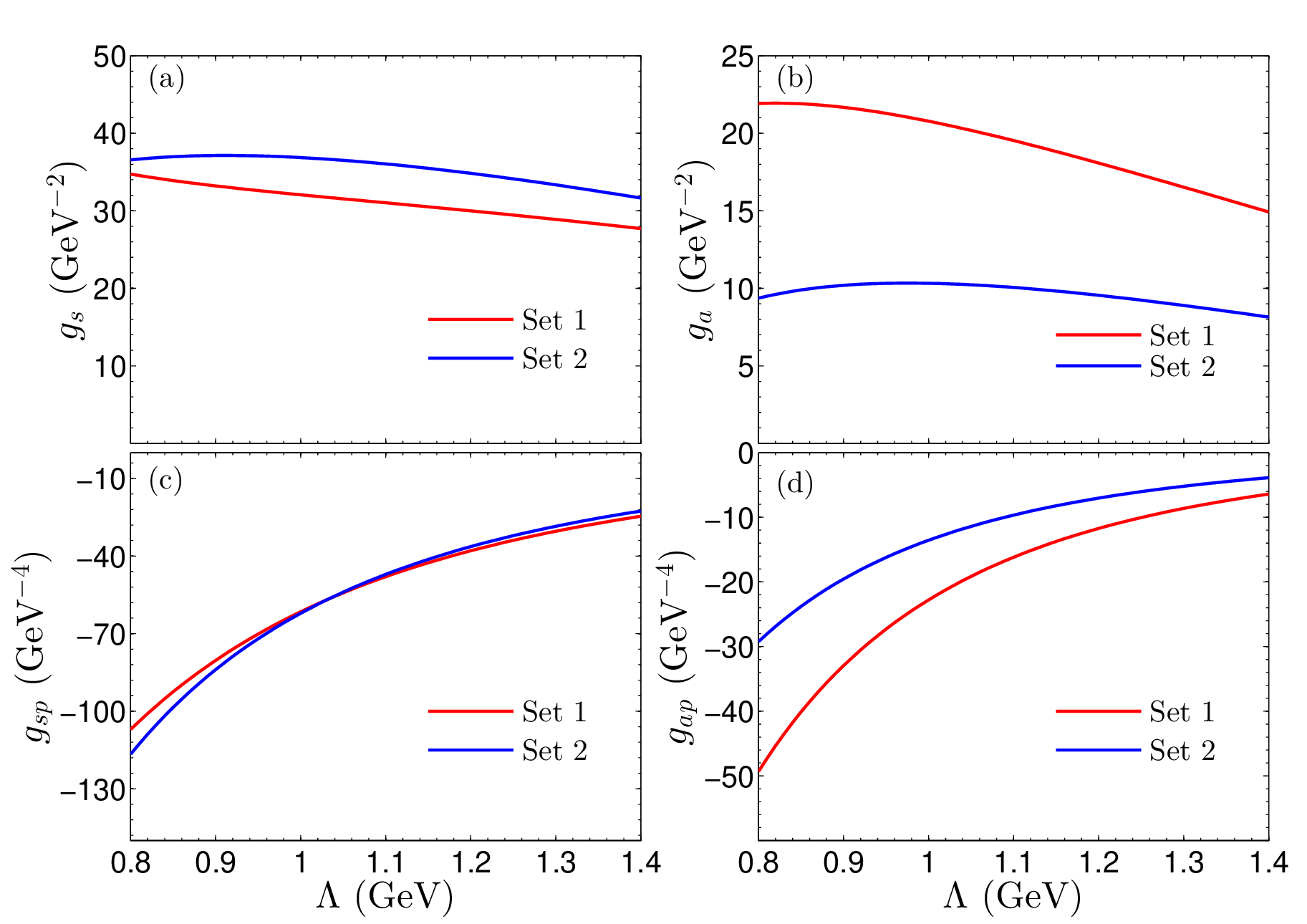}
    \caption{The $\Lambda$-dependences of the LECs ($g_{s}$, $g_{a}$, $g_{sp}$, $g_{ap}$) solved with the inputs from the Set 1 and 2. The red lines and blue lines denote the LEC solved from the Set 1 and 2, respectively.}
    \label{LECs}
\end{figure*}

\begin{figure*}[!htbp]
    \centering
    \includegraphics[width=0.9\linewidth]{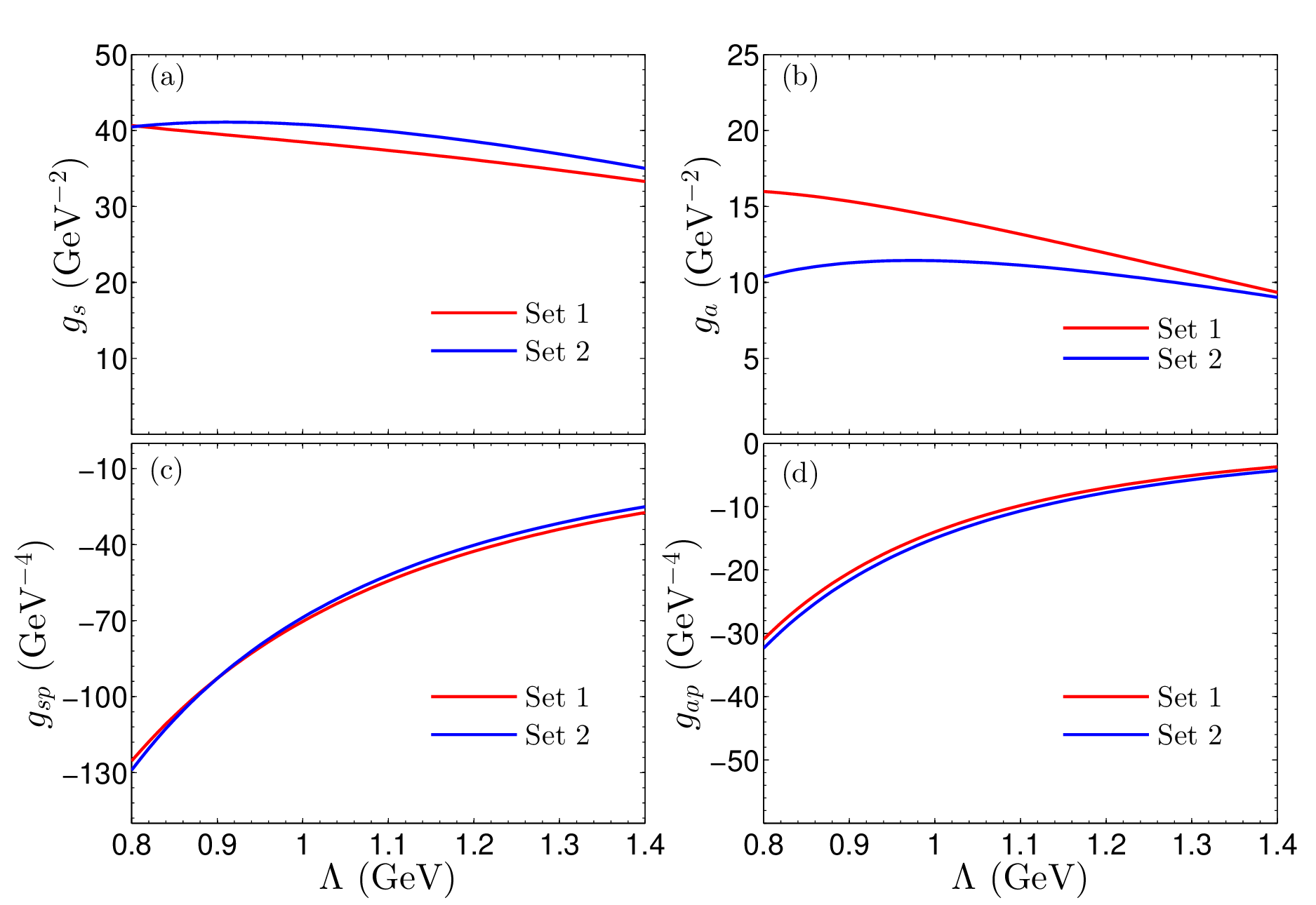}
    \caption{The $\Lambda$-dependences of the LECs ($g_{s}$, $g_{a}$, $g_{sp}$, $g_{ap}$) solved with the inputs from the Set 1 and 2. Instead of using the experimental width, we fix the width of the $Z_{cs}(4000)$ at 70 MeV in Set 1. The red lines and blue lines denote the LEC solved from the Set 1 and 2, respectively.}
    \label{LECs2}
\end{figure*}


We present the $\Lambda$ dependences of LECs extracted from the inputs of Set 1 and Set 2 in Fig. \ref{LECs}. As can be seen from Fig. \ref{LECs}, the LECs extracted from Set 1 and Set 2 have identical signs with comparable magnitudes. In the 0.8 GeV$\leq\Lambda\leq$1.4 GeV region, the parameters $\tilde{g}_s$, $\tilde{g}_a$, $\tilde{g}_{sp}$, and $\tilde{g}_{ap}$ have similar variation tendencies, this fact shows that the similarities of the LECs extracted from Set 1 and Set 2 have weak $\Lambda$ dependences.

The results presented in Fig. \ref{LECs} also shows that the obtained $\tilde{g}_a$ and $\tilde{g}_{ap}$ in Set 1 are different from that of Set 2. Here, if we have appropriately handled the SU(3) breaking effect among the $Z_c$ and $Z_{cs}$ states, then the differences of the $\tilde{g}_a$ and $\tilde{g}_{ap}$ in Set 1 and Set 2 mainly depend on the inputs of the central values of the selected $Z_c$ and $Z_{cs}$ states. At present, we only use the central values of the experimental data to extract the LECs form Set 1 and Set 2, and it is difficult to include the experimental uncertainties of the masses and widthes of the selected $Z_{cs}$ and $Z_c$ states into our analysis. The main reason is that if we include such uncertainties, the four LECs will also lie in the four solved regions, correspondingly. These solved regions may also extend in wide ranges, depending on the experimental uncertainties. Thus, comparing the four LECs wide ranges obtained from Set 1 and Set 2 can not give a significant similarity hint between the $Z_{cs}$ and $Z_c$ states.

Instead, we perform a numerical experiment, i.e., we adjust one of the experimental input, then we check if the similarities of the LECs extracted from the Set 1 and Set 2 can become better. We notice that the recent experiment from the BESIII collaboration reported the non-observation of the $Z_{cs}(3985)$ in the $J/\Psi K$ final states \cite{BESIII:2023wqy}. Besides, they fitted a small excess of $Z_{cs}$ over other components, the obtained mass and width are $4.044\pm0.006$ GeV and $0.036\pm0.016$, respectively. The significance of this small excess is only 2.3 $\sigma$. If such excess is related to a $Z_{cs}$ state, it may correspond to the $Z_{cs}(4000)$ reported from the LHCb collaboration \cite{LHCb:2021uow}. The resonance parameters of the $Z_{cs}(4000)$ from these two experiments are very different.

Here, we adopt the central value of the mass of $Z_{cs}(4000)$ from the LHCb, but treat the width of the $Z_{cs}(4000)$ as an adjustable parameter. We adjust the width of the $Z_{cs}(4000)$ and the SU(3) breaking factor $g_x$ to find the minimum $\chi$. We find that to obtain a minimum $\chi$, the $g_x$ is fixed at 1.27, and the width of the $Z_{cs}(4000)$ is adjusted to be 70 MeV. The results are presented in Fig. \ref{LECs2}, as can be seen from Fig. \ref{LECs2}, the LECs $\tilde{g}_s$, $\tilde{g}_a$, $\tilde{g}_{sp}$, and $\tilde{g}_{ap}$ extracted from the $Z_c$ and $Z_{cs}$ states (the width of the $Z_{cs}(4000)$ is fixed at 70 MeV) show very good consistences in a relatively big $\Lambda$ region. In this case, the resonance parameters of the observed $Z_{cs}$ and $Z_c$ states can be described simultaneously. In this framework, the differences of the effective potentials of $Z_{cs}$ and $Z_c$ states can be described by only introducing an SU(3) breaking factor $g_x$. This result inspires us to believe that the constructed framework might be a promising solution for a unified description of the $Z_{cs}$ and $Z_c$ states.

\section{Predictions to other $Z_{cs}$ and $Z_c$ states}\label{Sec 3D}
In this section, we give our predictions to the rest of $Z_{cs}$ and $Z_c$ states that are close to the thresholds of $D^{(*)}\bar{D}_s^{(*)}$ and $D^{(*)}\bar{D}^{(*)}$, respectively. We separately fit the LECs in the $Z_{cs}$ and $Z_c$ sectors with the inputs from Set 1 and Set 2 introduced in Sec. \ref{Sec 3C}. Each set consists of four quantities, the masses and widthes of the two $Z_{cs}$ or $Z_c$ states. Each mass or width includes three values, i.e., the experimental upper limit, central value, and lower limit. We consider different combinations of the three values of these four quantities to solve the corresponding LECs, and use the obtained LECs to calculate the lower and upper limits of the predicted $Z_{cs}$ or $Z_c$ state. The results are presented in Table \ref{Predictions}.
\begin{table*}[!htbp]
\renewcommand\arraystretch{1.5}
\setlength\tabcolsep{1.5pt} \caption{Our predictions to the possible $Z_c$ and $Z_{cs}$ resonances. We use the superscript ``$\dagger$'' to denote experimental inputs introduced in Set 1 and Set 2 for the $Z_{cs}$ and $Z_c$ states, respectively. All the results are in units of MeV.\label{Predictions}}
\begin{tabular}{cccc|cccccccc}
\toprule[1pt]
&\multicolumn{3}{c}{Our}&\multicolumn{4}{c}{Exp}\\
\hline
Threshold&State&Mass&Width&Threshold&State&Mass&Width&\\
\hline
3734.4&$|D\bar{D};0^{++}\rangle$&$-$&$-$
&3835.6&$|D\bar{D}_s;0^{+\tilde{+}}\rangle$&$3879.7\pm20.9$&$80.5\pm19.0$\\
3875.8&$|D\bar{D}^*;1^{++}\rangle$&$-$&$-$
&3979.3&$|D\bar{D}_s^*;1^{+\tilde{+}}\rangle$&${}^{\dagger}3985.2\pm2.6$&${}^{\dagger}13.8_{-7.2}^{+9.4}$\\
3875.8&$|D\bar{D}^*;1^{+-}\rangle$&${}^{\dagger}3887.1\pm2.6$&${}^{\dagger}28.6\pm2.6$
&3979.3&$|D\bar{D}_s^*;1^{+\tilde{-}}\rangle$&${}^{\dagger}4003.0^{+7.2}_{-15.2}$&${}^{\dagger}131.0\pm30.0$\\
4017.1&$|D^*\bar{D}^*;0^{++}\rangle$&${}^{\dagger}4051.0^{+24.0}_{-43.0}$&${}^{\dagger}82.0^{+52.0}_{-28.0}$
&4120.7&$|D^*\bar{D}_s^*;0^{+\tilde{+}}\rangle$&$4134.1\pm5.8$&$163.0\pm25.0$\\
4017.1&$|D^*\bar{D}^*;1^{+-}\rangle$&$4022.2\pm2.6$&$18.5\pm3.2$
&4120.7&$|D^*\bar{D}_s^*;1^{+\tilde{-}}\rangle$&$4142.9\pm7.0$&$125.4\pm44.6$\\
4017.1&$|D^*\bar{D}^*;2^{++}\rangle$&$-$&$-$
&4120.7&$|D^*\bar{D}_s^*;2^{+\tilde{+}}\rangle$&$4121.8\pm1.0$&$8.8\pm4.2$\\
\bottomrule[1pt]
\end{tabular}
\end{table*}

In Table \ref{Predictions}, we label the experimental inputs with superscript ``$\dagger$". As shown in Table \ref{Predictions}, in the $Z_c$ sector, except the input states in Set 2, we only find a $Z_{c}(4020)$ state, this state share identical effective potential to that of the $Z_c(3900)$ in the HQS limit, they have comparable $\delta_{M-M_{\text{Tr}}}$ values and widths. As listed in Table \ref{Exp}, the LHCb collaboration reported \cite{LHCb:2018oeg} a $Z_c(4100)$ in the $\eta_c\pi$ final states, the possible underlying structure of this state is still under debate \cite{Wang:2018ntv,Voloshin:2018vym,Zhao:2018xrd,Wu:2018xdi,Cao:2018vmv,Albuquerque:2018jkn,Sundu:2018nxt,Cao:2021ton,Baru:2021ddn,Yang:2021zhe}. The $J^{PC}$ number of this state could be $0^{++}$, if we consider the large uncertainties of the mass and width of $Z_c(4100)$, the $Z_c(4100)$ and $Z_c(4050)$ could be the same state. Besides, if we tentatively assign the $Z_{c}(4055)$ \cite{Belle:2014wyt} and $Z_{c}(4020)$ as the same state, then our framework could give an unified description of the observed $Z_c$ states (as listed in Table \ref{Exp}) that are close to the $D^{(*)}\bar{D}^{(*)}$ thresholds.

As shown in Table \ref{Predictions}, comparing to the $Z_c$ sector, in the $Z_{cs}$ sector, we obtain three extra $Z_{cs}$ states, i.e., the $|D\bar{D}_s;0^{+\tilde{+}}\rangle$, $|D\bar{D}_s^*;1^{+\tilde{+}}\rangle$, and $|D^*\bar{D}_s^*;2^{+\tilde{+}}\rangle$ states. Due to the SU(3) breaking effect, these three states do not have their SU(3) flavor $Z_c$ partners. We predict two $J^{PC}=0^{+\tilde{+}}$ $Z_{cs}$ states that are composed of the $D\bar{D}_s$ and $D^*\bar{D}^*_s$ components, they are all broad with widthes around 80 and 160 MeV, respectively. The $D\bar{D}_s$ and $\eta_c K$ are the possible decay channels for the $|D\bar{D}_s;0^{+\tilde{+}}\rangle$ state. Similarly, the $D\bar{D}_s$, $D^*\bar{D}_s^*$, and $\eta_c K$ are the possible decay channels for the $|D^*\bar{D}_s^*;0^{+\tilde{+}}\rangle$ state. We notice that the LHCb collaboration has measured the $\eta_c K$ invariant spectrum in the $B^0\rightarrow \eta_c K^+\pi^-$ process \cite{LHCb:2018oeg}. They found that without introducing some extra $Z_{cs}$ resonance contributions, it is possible to describe the $m(\eta_c K)$ and $m(K\pi)$ distribution well with the $K\pi$ contributions alone. However, we notice that there exist a dip at about 4050 MeV in the $\eta_c K$ invariant spectrum \cite{LHCb:2018oeg}, the obtained results lead us to conjecture that if such a dip could relate to the splitting of the predicted two $0^{+\tilde{+}}$ states. If these two states do exist, we also suggest to look for them in the invariant spectra of the $D\bar{D}_s$ and $D^*\bar{D}_s^*$ final states.

As shown in Table \ref{Predictions}, the ($|D\bar{D}_s^*;1^{+\tilde{+}}\rangle$, $|\bar{D}^*\bar{D}_s^*;2^{+\tilde{+}}\rangle$) and ($|D\bar{D}_s^*;1^{+\tilde{-}}\rangle$, $|\bar{D}^*\bar{D}_s^*;1^{+\tilde{-}}\rangle$) are the two pairs of the HQS partners. the states in each pair share comparable $\delta_{M-M_{\text{Tr}}}$ values and widthes. Among them, the predicted $|\bar{D}^*\bar{D}_s^*;1^{+\tilde{-}}\rangle$ may correspond to the $Z_{cs}(4220)$ if we consider the large experimental uncertainties of the $Z_{cs}(4220)$ from the LHCb collaboration \cite{LHCb:2021uow}. Besides, the predicted $|\bar{D}^*\bar{D}_s^*;2^{+\tilde{+}}\rangle$ state may correspond to the $Z_{cs}(4123)$ state reported from the BESIII collaboration \cite{BESIII:2022vxd}, this state has already been discussed in various models \cite{Du:2022jjv,Meng:2020ihj,Yang:2020nrt,Wang:2020rcx,Jin:2020yjn,Yan:2021tcp,Ortega:2021enc,Han:2022fup,Ikeno:2021mcb,Ding:2021igr,Giron:2021sla}, nevertheless, this state still needs further confirmation due to its low significance. Thus, we also give an unified description to the observed $Z_{cs}$ (as listed in Table \ref{Exp}) states that are close to the $D^{(*)}\bar{D}_s^{(*)}$ thresholds. Further measurements of the $Z_{cs}$ states will provide important inputs to our model, and will also provide important clues to test our theory.

\section{Summary}\label{Sec 4}
To summarise, in this work, we propose a possible framework to describe the observed $Z_c$ and $Z_{cs}$ states (listed in Table \ref{Exp}) that are close to the $D^{(*)}\bar{D}^{(*)}$ and $D^{(*)}\bar{D}_{s}^{(*)}$ thresholds, respectively.

We construct the effective potentials of the $Z_{cs}$ and $Z_c$ states by analogy with the effective potentials of the LO and NLO $N\bar{N}$ interactions. In the SU(3) flavor limit, according to the expressions of the effective potentials of the $N\bar{N}$ interactions, we reduce the LECs describing the effective potentials of the $Z_c$ and $Z_{cs}$ states into four parameters, i.e., $\tilde{g}_s$, $\tilde{g}_a$, $\tilde{g}_{sp}$, and $\tilde{g}_{ap}$. In addition, to identify the differences between the $Z_c$ and $Z_{cs}$ states, we further introduce an SU(3) breaking factor $g_x$, this factor is expect to be greater than 1 if we consider the different masses of the exchanged light mesons with different isospins.

Firstly, we determine the LECs $\tilde{g}_s$, $\tilde{g}_a$, $\tilde{g}_{sp}$, and $\tilde{g}_{ap}$ from the inputs of the experimental masses and widthes of the $Z_{cs}(4000)$ and $Z_{cs}(3985)$ states. They are assumed to be the $J^{P\tilde{C}}=1^{+\tilde{-}}$ and $1^{+\tilde{+}}$ states, respectively. Then we directly adopt the obtained four LECs to calculate the $J^{PC}=1^{+-}$ and $1^{++}$ states that are composed of the $D^{(*)}\bar{D}^*$. We run the undetermined parameter $g_x$ in the effective potentials of the $Z_c$ states and show that a considerable SU(3) breaking effect will lead the absences of the $|D\bar{D}^*; 1^{++}\rangle$ and $|D^*\bar{D}^*;2^{++}\rangle$ states, these two states should be the SU(3) partners of the $Z_{cs}(3985)$ and $Z_{cs}(4123)$ (the $Z_{cs}(4123)$ still need further confirmation), respectively. Besides, we also show that the SU(3) breaking effect will also reduce the width of the $|D\bar{D}^*; 1^{+-}\rangle$ state. This can qualitatively explain the large width difference between the $Z_c(3900)$ and $Z_{cs}(4000)$.

Then we compare the similarities between the $Z_{cs}$ and $Z_c$ states in another scheme. We determine the LECs from the inputs of the observed $Z_{cs}$ (Set 1) and $Z_{c}$ (Set 2) states separately. Then we compare the similarities of the LECs extracted from these two Sets. In this scheme, we fix the SU(3) breaking factor by finding the minimum $\chi$ at $\Lambda=1.0$ GeV. We show that the LECs obtained from the Set 1 and Set 2 are very close to each other, and this similarity has weak $\Lambda$ dependence. In particular, if we adjust the width of the mass of the $Z_{cs}(4000)$ to be 70 MeV, then the LECs extracted from Set 1 and Set 2 are almost the same, and have very weak $\Lambda$ dependences. This result lead us to believe that this framework might be a promising solution for a unified description of the $Z_c$ and $Z_{cs}$ states. Thus, further mearsurements on the resonance parameters of the observed $Z_c$ and $Z_{cs}$ states will provide important guidances to our calculation.

We also check the other possible resonances in the $Z_{cs}$ and $Z_c$ sectors. Comparing to the calculated $Z_c$ sector, the results in the $Z_{cs}$ sector may exist three extra $Z_{cs}$ states, i.e., the $|D\bar{D}_s;0^{+\tilde{+}}\rangle$, $|D\bar{D}_s^*;1^{+\tilde{+}}\rangle$, and $|D^*\bar{D}_s^*;2^{+\tilde{+}}\rangle$ states. The emergence of these three states is the consequence of the SU(3) breaking effect. Besides, we suggest to look for the $|D\bar{D}_s;0^{+\tilde{+}}\rangle$ state in the $D\bar{D}_s$ and $\eta_c K$ final states, and look for the $|D^*\bar{D}_s^*;0^{+\tilde{+}}\rangle$ state in the $D\bar{D}_s$, $D^*\bar{D}_s^*$, and $\eta_c K$ final states. With some reasonable assumptions, we place all the observed $Z_{cs}$ and $Z_c$ states into our framework, we hope that further explorations and measurements on these discussed $Z_{cs}$ and $Z_c$ states in the future can test our theory.
\section*{Acknowledgments}
Kan Chen want to thank Jian-Bo Cheng, Bo Wang, Lu Meng, and Prof. Shi-Lin Zhu and Xiang Liu for helpful discussion. Kan Chen is supported by the National Science Foundation of China under Grant No. 12305090. This project is supported by the National Science Foundation of China under Grant No. 12247103. 
This research is also supported by the National Science Foundation of China under Grants No. 11975033, No. 12070131001, and No. 12147168.

\end{document}